\documentstyle[preprint,aps]{revtex}
\begin{document}
\preprint{
SNUTP -98-104  $/\!/$
APCTP-98-013
}
\title{The Effects of the Gravitational Chern-Simons Term in the AdS$_{2+1}$ Geometry}
\author{Jin-Ho Cho
\thanks{e-mail address; jhcho@galatica.snu.ac.kr}
}
\address{Center for Theoretical Physics,
Seoul National University, Seoul, 151-742, Korea}
\address{APCTP,
207-43 Cheongryangri-dong, Dongdaemun-gu, Seoul,130-012, Korea}
\date{\today}
\maketitle
\begin{abstract}
We study the effect of the gravitational Chern-Simons term (GCST) in the (2+1)-dimensional anti-de Sitter (AdS$_{2+1}$) geometry. In the context of the gauge gravity, we obtain black hole solution and its boundary WZW theory. The BTZ black hole solution can still be retrieved but its gravitational mass and angular momentum become different from their inherent values. The deformation on these quantities due to the GCST can be summarized as $SO(1,1)$ times rescaling. The boundary WZW theory is found to be chiral, i.e., composed of the right moving part and the left moving part with different Kac-Moody levels. The statistical entropy is proportional to the area only for the large levels and vanishing GCST limit, but its coefficient is not the correct order in the Newton constant $G$. Some related physics are discussed.
\end{abstract}
\section{Introduction}

One of the most important successes of the D-brane physics is the statistical explanation of the black hole entropy in some specific cases \cite{strominger}. These include stringy black holes composed of BPS bound states of D-branes in $(4+1)$ \cite{callan} and $(3+1)$ dimensions \cite{malda1}\cite{johnson}. However these bound states correspond to the extremal black holes. Although the success can be continued to the near extremal cases \cite{horo}\cite{malda2}, it is still far from the general non-extremal cases including the Schwarzschild black hole.
  
One characteristic feature of the stringy black holes is that they interpolate between the asymptotic flat region and the near horizon adS geometry. Specifically these near horizon geometries are mostly related to the adS$_{2+1}$ via U-duality \cite{hyun}. On the other hand, Carlip's approach \cite{carlip1} to the statistical entropy for the BTZ black hole (black hole in adS$_{2+1}$)\cite{btz1} provides us with some hints about the holographic nature of the black hole. These facts together with the celebrated proposal for the adS/CFT duality \cite{maldacena} suggest that adS$_{2+1}$ geometry is the essential ingredient to the understanding of the statistical entropy for the general non-extremal black holes.
 
The $(2+1)$-dimensional gravity can be understood in the context of Chern-Simons gauge theory \cite{achucarro}\cite{witten}. Its gauge group is $SO(3,1)$, $ISO(2,1)$ or $SO(2,2)$ depending on the signature of the cosmological constant $\Lambda$ ($>0,\,\,=0$ or $\,<0$ respectively). As for the $ISO(2,1)$ case, the conical space and helical time structure appearing in the geometry of  Einstein gravity, for the massive spinning source, can also be seen in the $ISO(2,1)$ gauge gravity \cite{grignani}. However, the addition of the GCST makes difference between the Einstein gravity and the Gauge gravity. The resulting geometries for the massive spinning point source accord only in the asymptotic region \cite{cho}. 

One peculiar property of this asymptotic geometry is that the spin is induced by the mass via GCST. Our basic issue in this paper is to see whether this property is continued to the $SO(2,2)$ case and how the GCST deforms BTZ geometry and its boundary structure. Our analysis will be done in the context of gauge gravity. As for the case without the GCST, BTZ black hole solution is shown to be obtained in this context \cite{mansouri}. One important spirit here is that the geometry comes out of the massive spinning source. This is not clear in the Einstein gravity.

In the following section, we give some basic ingredients of the gauge gravity for the notational setup. In section \ref{sec3}, we show how the solution can be read from the holonomy generated by the source. We first get the solution for the gauge connection around the massive spinning point source. In order to map the nontrivial holonomy onto the `target' manifold, the Wilson loop operator is shown to be the identification Killing vector of  \cite{ban}. The black hole solution is constructed as the quotient adS$_{2+1}$.  In section \ref{sec4}, we study the boundary Wess-Zumino-Witten (WZW) theory that is deformed by the GCST. We obtain the relation between the conformal weight and the global charges. The central charges are also obtained and some related physics are discussed. In the final section, we conclude the paper with some remarks on the present status of BTZ black hole and suggest its future directions.

\section{Gravitational Chern-Simons Term in the $SO(2,2)$ Gauge gravity}

Let us start with some remarks on the Chern-Simons gauge gravity:
Differently from the $ISO(2,1)$ case, it is very confusing to consider the nontrivial black hole geometry in terms of the topological theory. In fact. the theory is defined on the `world' manifold $\Re\times\Sigma$. $\Re$ is parametrized by the `time' parameter $\tau$ and $\Sigma$ is two dimensional `spatial' hypersurface mostly taken as a disk parametrized by the `radial' coordinate $\rho$ and `polar' coordinate $\theta$. The geometry is constructed on the `target ' manifold via the Wilson loop operator which is the only observable in this topological theory. The `target' manifold is parametrized by the isovector in the fundamental representation of the corresponding gauge group. Therefore, the `world' manifold is fibered with the gauge connection and the isovector in the `target' manifold. The `target' manifold geometry can be pulled back to the `world' manifold via appropriately defined soldering form. 

The Einstein-Hilbert Lagrangian can be reproduced through the Chern-Simons Lagrangian.
\begin{eqnarray}
{\cal L}&=&-\frac{1}{16\pi G}<{\cal A}\,\,,\!\!\!\!\wedge d{\cal A}+\frac{2}{3}{\cal A}\wedge{\cal A}>\nonumber\\
&=&\frac{1}{16\pi G}({\cal R}+\frac{2}{l^2})\sqrt{-g}+\cdots.
\end{eqnarray}
This equation can be specified in more detail as follows. The Lie algebra $so(2,2)$ of the anti-de Sitter group reads as  
\begin{eqnarray}
[J_a,\,J_b]=\epsilon_{ab}{}^c J_c ,\,\,\,[P_a,\,J_b]=\epsilon_{ab}{}^c P_c ,\,\,\,[P_a,\,P_b]=\frac{1}{l^2}\epsilon_{ab}{}^c J_c,
\end{eqnarray}
where $P_a\equiv\frac{1}{l}J_{a3}$, $J_a\equiv -\frac{1}{2}\epsilon_{a}{}^{bc}J_{bc}$ and the Levi-Civita symbol is defined as $\epsilon^{012}=1$. The cosmological constant $\Lambda$ corresponds to $-1/l^2$. Since $SO(2,2)=SL(2,\Re{\sf })\times SL(2,\Re{\sf })$, making use of $J_a^{\pm}={1 \over 2}(J_a\pm l P_a)$ one can decouple the algebra as
\begin{eqnarray}
[ J_a^{\pm},J_b^{\pm} ]&=&\epsilon_{ab}{}^c J_c^{\pm},\,\,\,\,\,\,\,\,[J_a^+,J_b^-]=0
\end{eqnarray}
The bilinear form for this Lie algebra is generally determined as
\begin{eqnarray}
<J_a,\,J_b>=\alpha\eta_{ab},\,\,<P_a,\,J_b>=\eta_{ab},\,\,<P_a,\,P_b>=\frac{\alpha}{l^2}\eta_{ab} 
\end{eqnarray}
and in $SL(2,\Re{\sf })\times SL(2,\Re{\sf })$ basis as
\begin{eqnarray}
<J_a^{\pm},J_b^{\pm}>&=&{1 \over 2}(\alpha\pm l)\eta_{ab},\,\,\,<J_a^+,J_b^->=0.
\end{eqnarray}
Here, $\alpha$ is a parameter of the bilinear form and has the dimension of length.
With the $so(2,2)$ valued gauge connection one form
\begin{eqnarray}
{\cal A}&=&\omega^aJ_a+e^aP_a\nonumber\\
&=&(\omega^a+{e^a \over l})J_a^++(\omega-{e^a \over l})J_a^-\nonumber\\
&\equiv&A^aJ_a^++{\bar A}^aJ_a^-\equiv A+{\bar A},
\end{eqnarray}
we can expand the Chern-Simons Lagrangian in components.
\begin{eqnarray}
{\cal L}&=&-\frac{1}{\kappa}<{\cal A}\,\,,\!\!\!\!\wedge d{\cal A}+\frac{2}{3}{\cal A}\wedge{\cal A}>\nonumber\\
&=&-{1 \over \kappa}(e_a\wedge(2d\omega^a+\epsilon^a{}_{bc}\omega^b\wedge\omega^c)
+{1 \over {3l^2}}\epsilon_{abc}e^a\wedge e^b\wedge e^c\nonumber\\
&&+\alpha\omega_a\wedge(d\omega^a+{1 \over 3}\epsilon^a{}_{bc}\omega^b\omega^c)
+{\alpha\over l^2}e_a\wedge{\cal T}^a)\nonumber\\
&=&-{(l+\alpha) \over 2\kappa}A_a \wedge(d A^a+{1 \over 3}\epsilon^a{}_{bc}\,A^b\wedge A^c)\nonumber\\
&&+{(l-\alpha) \over 2\kappa}{\bar A}_{a}\wedge(d{\bar A}^a+{1 \over 3}\epsilon^a{}_{bc}\,{\bar A}^b\wedge{\bar A}^c),
\end{eqnarray}
where $\kappa\equiv 16\pi G$ and  ${\cal T}^a$ is the torsion two form which vanishes on shell in the source free region.

\section{Black Hole Geometry Read from the Holonomy}\label{sec3}
\subsection{ the Gauge Connection for the Point Source}
 
With the Chern-Simons action for the gauge fields:
\begin{eqnarray}\label{polar}
I_{CS}=-{(l+\alpha) \over {2\kappa}}\,\,Tr\int  A\,\wedge(dA+{2 \over 3}A\wedge A)+{(l-\alpha) \over {2\kappa}}\,\,Tr\int  {\bar A}\,\wedge(d{\bar A}+{2 \over 3}{\bar A}\wedge {\bar A}), \label{act1}
\end{eqnarray}
where $Tr(J^+_a J^+_b)=Tr(J^-_a J^-_b)=\eta_{ab}$ and $Tr(J^+_a J^-_b)=0$, we consider the action for the source:
\begin{eqnarray}\label{source}
I_{S}=\int d\tau(\Pi_A {\dot q}^A+A_{a\tau}{\cal J}^a+{\bar A}_{a\tau}{\bar {\cal J}}^a+\zeta(q^Aq_A+l^2)+\zeta_+({\cal J}^a{\cal J}_a-j^2)+\zeta_-({\bar {\cal J}}^a{\bar {\cal J}}_a-{\bar j}^2),
\end{eqnarray}
where $q^A (A=0,1,2,3)$ are the isovector components specifying the anti-de Sitter coordinates of the source on the hypersurface $q^Aq_A+l^2=0$ and $\Pi_A$ are their conjugates. The index contraction is done with respect to the flat metric $\eta_{AB}=diag(-,+,+,-)$ of the embedding space. The spin currents ($SL(2,\Re{\sf })\times SL(2,\Re{\sf })$ currents in the field representation) ${\cal J}^a$, ${\bar {\cal J}}^a$ are constrained by the on-shell conditions ${\cal J}^a{\cal J}_a-j^2=0$, ${\bar{\cal J}}^a{\bar{\cal J}}_a-j^2=0$ respectively. Here, $j^2$ and ${\bar j}^2$ are the two Casimir invariants of the $SO(2,2)$ group, which will be specified further below.

The equations of motion can be derived as
\begin{eqnarray}\label{eom}
&&\partial_iA^a_0-\partial_0A^a_i+\epsilon^a{}_{bc}A^b_iA^c_0=0,\nonumber\\
&&\partial_i{\bar A}^a_0-\partial_0{\bar A}^a_i+\epsilon^a{}_{bc}{\bar A}^b_i{\bar A}^c_0=0,
\end{eqnarray}
with the Gaussian constraints: 
\begin{eqnarray}
&&\epsilon^{ij}F^a_{ij}={{2\kappa} \over (\alpha+l)}{\cal J}^a\delta^{(2)}(x-x_0),\nonumber\\
&&\epsilon^{ij}\bar{F}^a_{ij}={{2\kappa} \over (\alpha-l)}{\bar{\cal J}}^a\delta^{(2)}(x-x_0).
\end{eqnarray}
where the curvature is defined by $F^a_{ij}\equiv\partial_iA^a_j-\partial_jA^a_i+\epsilon^a{}_{bc}A^b_iA^c_j$ and $\bar{F}^a_{ij}\equiv\partial_i{\bar A}^a_j-\partial_j{\bar A}^a_i+\epsilon^a{}_{bc}{\bar A}^b_i{\bar A}^c_j$.

The parametrization of the anti-de Sitter hypersurface $q^Aq_A+l^2=0$ can be given in many ways, some conventional ones of which are well illustrated in the appendix of \cite{kraus}. The most relevant one in this paper is the BTZ coordinates.
For $0\leq\hat{r}\leq l$,
\begin{eqnarray}
q^0=\pm\hat{r}\cosh{\hat{\phi}},\,\,\,q^1=\hat{r}\sinh{\hat{\phi}},\,\,\,
q^2=\sqrt{l^2-\hat{r}^2}\sinh{\hat{t}} ,\,\,\,q^3=\pm\sqrt{l^2-\hat{r}^2}\cosh{\hat{t}}, 
\end{eqnarray}
while for $l\leq\hat{r}<\infty$,
\begin{eqnarray}
q^0=\pm\hat{r}\cosh{\hat{\phi}},\,\,\,q^1=\hat{r}\sinh{\hat{\phi}},\,\,\,
q^2=\pm\sqrt{\hat{r}^2-l^2}\cosh{\hat{t}} ,\,\,\,q^3=\sqrt{\hat{r}^2-l^2}\sinh{\hat{t}}, 
\end{eqnarray}
In both coordinate patches, the metric can be written down as 
\begin{eqnarray}
ds^2=-(dq^0)^2+(dq^1)^2+(dq^2)^2-(dq^3)^2=-(\hat{r}^2-l^2)d\hat{t}^2+{l^2 \over (\hat{r}^2-l^2)}d\hat{r}^2+\hat{r}^2d\hat{\phi}^2.\label{metric}
\end{eqnarray}

The point source is at $\rho=0$ on the world volume.  Where should it be on the `target' manifold, i.e., adS space? Since the BTZ coordinates do not cover the whole manifold of  adS space, it is quite difficult to pinpoint the exact image of $\rho=0$ point on the BTZ coordinate patches. However to achieve the rotationally symmetric solution in the BTZ coordinates, it is appropriate to assume that it is located behind $\hat{r}=0$ point, that is to say, somewhere satisfying $\hat{r}^2\leq 0$. This will be justified in the solution later. Therefore in the BTZ coordinates, its worldline is spacelike and so are their currents. In order to achieve the constraints ${\cal J}^a{\cal J}^b\eta_{ab}=j^2$ and ${\bar{\cal J}}^a{\bar{\cal J}}^b\eta_{ab}={\bar j}^2$, we set ${\cal J}^a=(0,0,j)$ and $\bar{\cal J}^a=(0,0,\bar{j}).$ With an appropriate choice of the gauge fixing conditions, $A^a_\rho=\bar{A}^a_\rho=0$, the relevant non-vanishing components are
\begin{eqnarray}\label{sol}
A^2_\theta={{8Gj} \over{\alpha+l}},\,\,\,{\bar A}^2_\theta={{8G{\bar j}} \over{\alpha-l}}.
\end{eqnarray}

One might think this solution results in degenerate metric due to the vanishing components $e^a_{\rho}$. However it should be noted that the components $e^a_{\mu}$ are no longer the soldering form in the gauge gravity. In fact, the composition $e^a_{\mu}e^a_{\nu}\eta_{ab}$ is not gauge invariant quantity. The correct soldering form can  be defined as the covariant derivative of the anti-de Sitter coordinates ${\cal E}^A_{\mu}={\cal D}_\mu q^A$. 

Therefore in order to obtain the metric $g_{\mu\nu}={\cal E}^A_{\mu}{\cal E}^B_{\nu}\eta_{AB}$, one must know the precise dependency of the anti-de Sitter coordinates $q^A$ on the `world' coordinates $(\tau,\,\rho,\,\theta)$, which is somewhat cumbersome job in this case where $q^A$ are the embedding coordinates of the anti-de Sitter hypersurface rather than the coordinates of the anti-de Sitter space itself (while in the Poincar\'{e} case, this can be easily done \cite{cho}). 

\subsection{Wilson Loop Operator as the Identification Killing Vector}

In the Chern-Simons gauge gravity, one can take rather a different route, i. e., reparametrize the anti-de Sitter hypersurface with the image  $(t(\tau),\,r(\rho),\,\phi(\theta))$ of the `world' coordinates so that the holonomy structure is well incorporated. This means to find the dependency of the BTZ coordinates $(\hat{t},\,\hat{r},\,\hat{\phi})$ on the image coordinates $(t(\tau),\,r(\rho),\,\phi(\theta))$.
Once this is done, one can construct the metric in terms of the coordinates $(t,\,r,\,\phi)$ in the same way as (\ref{metric}). 

The former method is well summarized in \cite{grig} and is worked out in \cite{cho} for the Poincar\'{e} gravity with the GCST. The latter method is adopted in \cite{vaz} for the Poincar\'{e} case and in the recent paper \cite{mansouri} for the anti-de Sitter case without the GCST.
In this section, we follow this latter method for the anti-de Sitter case with the GCST. The scheme is very close to that of  \cite{mansouri}, but differently from which we work in the BTZ coordinates rather than the global coordinates \cite{kraus}.

The two Casimir invariants of $SO(2,2)$ can be defined as
\begin{eqnarray}
(J^+-J^-)^2\equiv{r_+{}^2 \over 16G^2},\,\,\,(J^++J^-)^2\equiv{r_-{}^2 \over 16G^2}.
\end{eqnarray}
Equivalently one can introduce another set of invariants $M$ and $J$ (not to be confused with Lie algebra generators $J$'s),
\begin{eqnarray}\label{mj}
r_+{}^2+r_-{}^2\equiv 8GMl^2,\,\,\,r_+r_- \equiv 4GJl 
\end{eqnarray}

The solution (\ref{sol}) can be written in terms of these Casimir invariants as
\begin{eqnarray}
\omega^2{}_\theta&=&{4G \over {\alpha^2-l^2}}(\alpha(j+\bar{j})-l(j-\bar{j}))={\alpha r_--lr_+\over {\alpha^2-l^2}}\equiv{R_+ \over l},\nonumber\\
{e^2{}_\theta\over l}&=&{4G \over {\alpha^2-l^2}}(\alpha(j-\bar{j})-l(j-\bar{j}))={\alpha r_+-lr_-\over {\alpha^2-l^2}}.\equiv{R_- \over l}.
\end{eqnarray}
The corresponding nontrivial holonomy can be summarized as the Wilson loop operators:
\begin{eqnarray}
{\cal W}[\omega]={\cal P}\exp{\oint{\omega^2{}_\theta}J_2},\,\,\,{\cal W}[e]={\cal P}\exp{\oint{e^2{}_\theta}P_2},\label{wil}
\end{eqnarray}
where ${\cal P}$ denotes the path ordered product. With the representation $J_2=-J_{01}=\pmatrix{
0 & 1 \cr
1 & 0 \cr
}$ on $(q^0,\,q^1)$ and $lP_2=J_{23}=\pmatrix{
0 & 1 \cr
1 & 0 \cr
}$ on $(q^2,\,q^3)$, the Wilson line operators become
\begin{eqnarray}\label{wilson}
{\cal W}_\phi[\omega]&=&I\cosh{\omega^2{}_\theta\phi}+J_2\sinh{\omega^2{}_\theta\phi}\nonumber\\
{\cal W}_\phi[e]&=&I\cosh{{e^2{}_\theta\over l}\phi}+lP_2\sinh{{e^2{}_\theta\over l}\phi},
\end{eqnarray}
where $\phi$ is the image parameter of  $\theta$, on the `target' space. In fact, these are the boosting operators in the $(q^0,q^1)$-plane and $(q^2,q^3)$-plane respectively. For $\phi=2\pi n$, these operators make multiple images of $\theta=0\simeq2\pi$, on the $(q^0,q^1)$-plane and $(q^2,q^3)$-plane.

We are to encode the holonomy property of the `world'manifold onto the `target' manifold by identifying those multiple images. The gauge connection components ${\cal A}_\theta=\omega^2{}_\theta J_2+e^2{}_\theta P_2$  in the exponent of the Wilson loop operator is nothing but the identification Killing vector $\xi$ of \cite{ban}, which will be clearer below.
\begin{eqnarray}
\xi&=&{1 \over \alpha^2-l^2}\left((\alpha r_--lr_+)J_2+(\alpha r_+-lr_-)lP_2\right)\nonumber\\
&=&{R_+ \over l}(q^1\partial_0+q^0\partial_1)+{R_- \over l}(q^2\partial_3+q^3\partial_2),
\end{eqnarray}
which is spacelike in the region
\begin{eqnarray}\label{border}
&&-{q^0}^2+{q^1}^2\leq{l^2(\alpha r_+-lr_-)^2 \over (l^2-\alpha^2)(r_+^2-r_-^2)}\nonumber\\
&&-{q^2}^2+{q^3}^2\leq{l^2(\alpha r_--lr_+)^2 \over (l^2-\alpha^2)(r_+^2-r_-^2)},
\end{eqnarray}
which means $\hat{r}^2\geq-{l^2(\alpha r_+-lr_-)^2 \over (l^2-\alpha^2)(r_+^2-r_-^2)}$. Therefore in the whole region of BTZ coordinate patches, the Killing vector $\xi$ remains spacelike.

For better understanding, one can rewrite the Killing vector in the BTZ coordinates. The only thing to be done is to represent $J_2$ and $P_2$ in terms of  BTZ coordinates. Those derivatives in the anti-de Sitter coordinates are translated as 
\begin{eqnarray}
\partial_0=\cosh{\hat{\phi}}{\partial \over \partial\hat{r}}
-{\sinh{\hat{\phi}}\over \hat{r}}{\partial \over \partial\hat{\phi}},\,\,\,
\partial_1=-\sinh{\hat{\phi}}{\partial \over \partial\hat{r}}
+{\cosh{\hat{\phi}}\over \hat{r}}{\partial \over \partial\hat{\phi}},
\end{eqnarray}
which is valid for the whole region.
In the region $l\leq\hat{r}<\infty$,
\begin{eqnarray}
\partial_2={\sqrt{\hat{r}^2-l^2}  \over \hat{r}}\cosh{\hat{t}}{\partial \over \partial\hat{r}}
-{\sinh{\hat{t}}\over \sqrt{\hat{r}^2-l^2} }{\partial \over \partial\hat{t}},\,\,\,
\partial_3=-{\sqrt{\hat{r}^2-l^2}  \over \hat{r}}\sinh{\hat{t}}{\partial \over \partial\hat{r}}
+{\cosh{\hat{t}}\over \sqrt{\hat{r}^2-l^2}}{\partial \over \partial\hat{t}},
\end{eqnarray}
while in the region $0\leq\hat{r}\leq l$,
\begin{eqnarray}
\partial_2={\sqrt{l^2-\hat{r}^2}  \over \hat{r}}\sinh{\hat{t}}{\partial \over \partial\hat{r}}
+{\cosh{\hat{t}}\over \sqrt{l^2-\hat{r}^2} }{\partial \over \partial\hat{t}},\,\,\,
\partial_3=-{\sqrt{l^2-\hat{r}^2}  \over \hat{r}}\cosh{\hat{t}}{\partial \over \partial\hat{r}}
-{\sinh{\hat{t}}\over \sqrt{l^2-\hat{r}^2}}{\partial \over \partial\hat{t}}.
\end{eqnarray}
In both regions, the following descriptions are valid.
\begin{eqnarray}
q^1\partial_0+q^0\partial_1={\partial \over \partial\hat{\phi}},\,\,\,
q^2\partial_3+q^3\partial_2={\partial \over \partial\hat{t}}. 
\end{eqnarray}

The Killing vector $\xi$ is finally given in the BTZ coordinates as
\begin{eqnarray}
\xi={1\over l^2-\alpha^2}\left( (lr_+-\alpha r_-){\partial \over \partial\hat{\phi}}
+(lr_--\alpha r_+){\partial \over \partial\hat{t}} \right)
 ={R_+ \over l}{ \partial\over \partial \hat{\phi}}+{R_- \over l}{ \partial\over \partial\hat{t}},
\end{eqnarray}
which is nothing but the linear combination of those two boosting generators in the $(q^0,q^1)$-plane and $(q^2,q^3)$-plane. (Here, it becomes transparent that the Wilson loop operators correspond to the identification Killing vector of \cite{ban}.)
Therefore, encircling the source once gives the transformation specified by the Wilson loop operators  (\ref{wil}), according to which $\hat{\phi}$ and $\hat{t}$ coordinates are translated by ${2\pi R_+\over l} $ and ${2\pi R_-\over l} $ respectively. We introduce new coordinates which make these holonomy properties most explicit.
\begin{eqnarray}
\hat{\phi}(\phi,t)&=&{lr_+-\alpha r_-\over l^2-\alpha^2} \phi+\cdots\nonumber\\
\hat{t}(\phi,t)&=&{lr_--\alpha r_+ \over l^2-\alpha^2} \phi+\cdots,
\end{eqnarray}
where $\phi$ is just the image parameter of $\theta$, appearing in (\ref{wilson}). In order for this transformation to be nondegenerate, at  least  one of  the `$\cdots$' terms must  have $t$ dependency. However in any case, we can make these coordinate transformations be more symmetrical in $\phi$ and $t$. The result is as follows.
\begin{eqnarray}
\left (\matrix{
\hat{t}(\phi,t)\cr
\hat{\phi}(\phi,t)\cr
}\right )={1 \over l^2-\alpha^2}\pmatrix{
(lr_+-\alpha r_-) & (lr_--\alpha r_+) \cr
(lr_--\alpha r_+) & (lr_+-\alpha r_-) \cr
}\left (\matrix{
{t \over l}\cr
\phi\cr
}\right )=\pmatrix{
{R_+ \over l} & {R_- \over l} \cr
{R_- \over l} & {R_+ \over l} \cr
}\left (\matrix{
{t \over l}\cr
\phi\cr
}\right ).\label{transf}
\end{eqnarray}
In fact, this is the very transformation which makes $\xi$ be exactly along the spatial direction. The Killing vector becomes simplified as $\xi={\partial \over \partial\phi}$ in the whole region of  BTZ coordinates.

A few thing remarkable is that the transformation (\ref{transf}) can be considered as the combination of the rescaling and Lorentz boosting:
\begin{eqnarray} 
\exp{\lambda}\cdot\pmatrix{
\cosh{\gamma} & \sinh{\gamma} \cr
\sinh{\gamma} & \cosh{\gamma} \cr
},
\end{eqnarray}
 where $\exp{\lambda}={ \sqrt{r_+^2-r_-^2}  \over \sqrt{l^2-\alpha^2} }$ and $\cosh{\gamma}={lr_+-\alpha r_- \over \sqrt{(l^2-\alpha^2)(r_+^2-r_-^2)}}$ and $\sinh{\gamma}={lr_--\alpha r_+ \over \sqrt{(l^2-\alpha^2)(r_+^2-r_-^2)}}$. In the new coordinates, the metric (\ref{metric}) becomes
\begin{eqnarray}
ds^2&=&\exp{2\lambda}\left(- {\hat{r}^2(\hat{r}^2-l^2) \over \hat{r}^2+l^2\sinh^2{\gamma}}{dt^2 \over l^2} 
+(\hat{r}^2+l^2\sinh^2{\gamma})
(d\phi+{l^2\cosh{\gamma}\sinh{\gamma} \over \hat{r}^2+l^2\sinh^2{\gamma}}{dt \over l})^2\right)\nonumber\\
&&+{l^2 \over \hat{r}^2-l^2}d\hat{r}^2.
\end{eqnarray}
As is noted above, the holonomy gives translation along $\phi$ by $2\pi$. 

\subsection{Black Hole Solution as the Quotient AdS$_{2+1}$}

It becomes natural to compactify the $\phi$ direction like $\phi\sim\phi+2\pi$, to encode the holonomy property of the `world' manifold onto the `target' manifold, then  we get the black hole solution. The above metric can be cast into the standard form via further coordinate transformation setting $\exp{2\lambda}\cdot(\hat{r}^2+l^2\sinh{^2\gamma})\equiv r^2$. In this new radial coordinate, the border line of (\ref{border}) which ensures the spacelikeness of the identification Killing vector $\xi$ corresponds to $r=0$ `point'. The final form of the metric is
\begin{eqnarray}\label{bh}
ds^2&=&-{(r^2-l^2\exp{2\lambda}\cosh^2{\gamma})
(r^2-l^2\exp{2\lambda}\sinh^2{\gamma}) \over r^2}{dt^2 \over l^2}\nonumber\\
&&+r^2(d\phi+{l^2\exp{2\lambda}\cosh{\gamma}\sinh{\gamma} \over r^2}{dt \over l})^2\nonumber\\
&&+{l^2 r^2 \over (r^2-l^2\exp{2\lambda}\cosh^2{\gamma})
(r^2-l^2\exp{2\lambda}\sinh^2{\gamma})}dr^2.
\end{eqnarray}
This is nothing but the BTZ black hole. The only difference from the BTZ solution is that the radii $r_+,\,r_-$ of the two horizons have been deformed to the effective values $R_+,\,R_-$.
\begin{eqnarray}\label{radius}
\left (\matrix{
R_+\cr
R_-\cr
}\right )
={l \over l^2-\alpha^2}\pmatrix{
l & -\alpha \cr
-\alpha & l \cr
}
\left (\matrix{
r_+\cr
r_-\cr
}\right )
\end{eqnarray}

\subsection{Deformed BTZ Spectrum}

Making use of the same relation as (\ref{mj}), one can equivalently define the gravitational angular momentum $\bar{J}$ and the gravitational mass $\bar{M}$, which can be summarized as 
\begin{eqnarray}\label{MJ}
\left (\matrix{
\bar{M}l\cr
\bar{J}\cr
}\right )={l^2 \over (l^2-\alpha^2)^2}\pmatrix{
l^2+\alpha^2 & -2\alpha l\cr
-2\alpha l & l^2+\alpha^2 \cr
}\left (\matrix{
Ml\cr
J\cr
}\right ).
\end{eqnarray}
Interestingly, the transformations eq. (\ref{radius}) and eq. (\ref{MJ}) of  these two sets of the second Casimir can be considered as the combination of the rescaling and Lorentz boosting. Specifically in (\ref{MJ}), these transformations do not alter the signature of $M^2l^2-J^2$. Therefore, extremal black hole remains extremal. As the GCST turned on, the gravitational mass and angular momentum become different from their corresponding inherent partners. The adS space ($M=-{1 \over 8G},\,J=0$) attains the naked singular point while the naked singular black hole ($M=-{1 \over 8G}{l^2+\alpha^2 \over l^2},\,J=-{1 \over 8G}{2\alpha \over l}$) becomes the adS space. As for BTZ black hole spectrum, one observes that  the angular momentum can be induced by the mass and the mass can be induced by the spin. This is very similar to the Poincar\'{e} case \cite{cho}. Indeed as one makes the cosmological constant vanishingly small (limit $l\rightarrow\infty$), the gravitational angular momentum  reads as $\bar{J}\sim -2\alpha M$, even when the inherent value is zero. (In the Poincar\'{e} case, the relation is $\bar{J}=\alpha M$.) 

It is also interesting to see that the transformation (\ref{transf}) can be decomposed into the form
\begin{eqnarray}\label{decompose}
\left (\matrix{
\hat{t}\cr
\hat{\phi}\cr
}\right )&=&{1 \over l^2-\alpha^2}\pmatrix{
l & -\alpha \cr
-\alpha & l \cr
}\pmatrix{
r_+ & r_- \cr
r_- & r_+ \cr
}\left (\matrix{
{t \over l}\cr
\phi\cr
}\right )\nonumber\\
&=&\exp{\lambda}\cdot\pmatrix{
\cosh{\gamma_1} & \sinh{\gamma_1} \cr
\sinh{\gamma_1} & \cosh{\gamma_1} \cr
}\pmatrix{
\cosh{\gamma_2} & \sinh{\gamma_2} \cr
\sinh{\gamma_2} & \cosh{\gamma_2} \cr
}\left (\matrix{
{t \over l}\cr
\phi\cr
}\right ),
\end{eqnarray}
where  $\cosh{\gamma_1}={l \over \sqrt{l^2-\alpha^2} } $  and
$\cosh{\gamma_2={r_+ \over \sqrt{r_+^2-r_-^2} }}$. Therefore the second transformation is just the  Lorentz boosting along the $\hat{\phi}$-direction which makes, in the $\alpha=0$ case, the identification Killing vector $\xi$ be in accord to the $\phi$ direction. This factor becomes infinite boosting in the extreme limit $r_+\sim r_-$. The first boosting is the effect of deformation due to the GCST. Here one can see the kinematics of the induced angular momentum. The GCST gives the boosting along the $\hat{\phi}$-direction. This means the system is in the rotating frame, which effectively generates the angular momentum on the geometry.   
The case without the gravitational Chern-Simons term can be obtained by setting  $\alpha=0$.

\section{Boundary WZW Theory}\label{sec4}

One characteristic feature of the AdS space is its timelike boundary. Specifically as for the Chern-Simons formulation of $(2+1)$-dimensional gravity, the presence of this boundary partially breaks the gauge symmetry through the boundary term to induce the boundary WZW theory\cite{carlip1}\cite{ban1}. This section deals with this boundary theory and its symmetry. For that purpose, it is convenient to rewrite those `right' and `left' Chern-Simons term in (\ref{act1}) as follows.
\begin{eqnarray}
I_{CS}=-{k \over 4\pi}\,\,Tr\int_{\cal M}  A\,\wedge(dA+{2 \over 3}A\wedge A)+{\bar{k} \over 4\pi}\,\,Tr\int_{\cal M}  {\bar A}\,\wedge(d{\bar A}+{2 \over 3}{\bar A}\wedge {\bar A}), \label{chiral}
\end{eqnarray}
where $k={2\pi(l+\alpha) \over \kappa}$ and $\bar{k}={2\pi(l-\alpha) \over \kappa}$.

The precise form of the boundary term depends on the boundary condition. Therefore the next step to do is to determine the boundary condition. We take $A_\tau+A_\theta\equiv A_v=0$ and $\bar{A}_\tau-\bar{A}_\theta\equiv A_u=0$ as the asymptotic boundary condition for the BTZ black hole ($v=(\tau+\theta)/2,\,\,u=(\tau-\theta)/2$). One can see {\it a posteriori}, these are the natural candidates to be incorporated into the boundary conformal structure. Indeed inserting these conditions into the equations of motion (\ref{eom}) results in $\partial_vA_\theta=(\partial_\tau+\partial_\theta)A_\theta=0$ and $\partial_u\bar{A}_\theta=(\partial_\tau-\partial_\theta)\bar{A}_\theta=0$ respectively. Therefore $A_\theta$ is right moving and $\bar{A}_\theta$ is left moving with the light speed. This sounds plausible because only the massless mode can be viable in the asymptotic region. 
In the metric formulation, an equivalent boundary condition was used; the geometry is asymptotically anti-de Sitter space in \cite{brown}.   

The appropriate boundary terms corresponding to these conditions will be 
\begin{eqnarray}
I_B=-{k \over 4\pi}Tr\int_{\partial{\cal M}} A_v\,A_\theta 
+{\bar{k} \over 4\pi}Tr\int_{\partial{\cal M}}\bar{A}_u\,\bar{A}_\theta.\label{chiralb}
\end{eqnarray}
This boundary action breaks the gauge symmetry partially, due to which some portion of the gauge degrees become converted into physical degrees. In order to extract these physical degrees, we perform the gauge variation for the total action, i.e., the equation (\ref{chiral}) plus (\ref{chiralb}) together with the source part (\ref{source}). However, the source part is irrelevant to the outer boundary theory. It only contributes to the Gauss law constraints which might be important in our analysis of the global charge. Therefore we restrict our attention to the two parts $I_{CS}+I_B$ with the notion that the Gauss law constraints should be supplemented by the source term.
\begin{eqnarray}\label{split}
(I_{CS}+I_B)[A,\bar{A};g,\bar{g}]&=&(I_{CS}+I_B)[A,\bar{A}]\nonumber\\
&&-{k \over 2\pi}Tr\int_{\partial{\cal M}} g^{-1}\partial_vg\,\, g^{-1}A_\theta g
-{k\over 4\pi}Tr\int_{\partial{\cal M}} g^{-1}\partial_vg\,\, g^{-1}\partial_\theta g\nonumber\\
&&+{k \over 12\pi}Tr\int g^{-1}dg\wedge g^{-1}dg\wedge g^{-1}dg
\nonumber\\
&&+{\bar{k} \over 2\pi}Tr\int_{\partial{\cal M}}
\bar{g}^{-1}\partial_u\bar{g}\,\,\bar{g}^{-1}\bar{A}_\theta \bar{g}
+{\bar{k}\over 4\pi}Tr\int_{\partial{\cal M}}
\bar{g}^{-1}\partial_u\bar{g}\,\,\bar{g}^{-1}\partial_\theta \bar{g}\nonumber\\
&&-{\bar{k} \over 12\pi}Tr\int \bar{g}^{-1}d\bar{g}\wedge \bar{g}^{-1}d\bar{g}\wedge \bar{g}^{-1}d\bar{g}
\end{eqnarray}
This can be understood as the split of the boundary degrees of the freedom $g^{-1}dg$ from the bulk degrees of the freedom $A,\,\bar{A}$. The action (\ref{split}) describes their interaction and dynamics. Essentially the action exhibits no difference from the case without the GCST. The only difference is the `unmatched' levels $k$ and $\bar{k}$.

We are to study the asymptotic symmetry for the system (\ref{split}). There are a lot of ways for this. One can explicitly work out the symplectic structure of the boundary theory and find the algebra for the Noether charge \cite{lee}. This might be the essential step to pursue if one are to get the precise boundary action. Another smart way is the Regge-Teitelboim method \cite{regge}, which gives the interpretation for the global charge as the generators of the residual gauge group after the gauge is fixed. \cite{ban1} studies the global charges for the BTZ black hole geometry. One strong point of this latter method is that one needs not know the precise form of the boundary action. We follow this latter method for brevity.

Making use of the Poisson algebra
\begin{eqnarray}
\{A^a_i,\,A^b_j\}=-{2\pi \over k}\eta^{ab}\epsilon_{ij},\,\,\,\{\bar{A}^a_i,\,\bar{A}^b_j\}={2\pi \over \bar{k}}\eta^{ab}\epsilon_{ij},
\end{eqnarray}
one can show the constraints $G^a\equiv-{k \over 4\pi}\epsilon^{ij}F^a_{ij}+{\cal J}^a$ and $\bar{G}^a\equiv{\bar{k} \over 4\pi}\epsilon^{ij}\bar{F}^a_{ij}+\bar{{\cal J}}^a$ satisfy the standard algebra of $SO(2,1)\times SO(2,1)$ (note that the anti-de Sitter currents ${\cal J}$and $\bar{\cal J}$ are expected to satisfy the same algebra),
\begin{eqnarray}
\{G^a,\,G^b\}=f^{ab}{}_cG^c,\,\,\,\,\,\{\bar{G}^a,\,\bar{G}^b\}=f^{ab}{}_c\bar{G}^c.
\end{eqnarray}
The global charges $Q(\eta)$ and $\bar{Q}(\bar{\eta})$ are determined from the differentiability requirement for the smeared operators,
\begin{eqnarray}
G(\eta)=\int_{\Sigma}\eta^aG_a+Q(\eta)\approx Q(\eta),\,\,\,\,\,\bar{G}(\bar{\eta})=\int_{\Sigma}\bar{\eta}^a\bar{G}_a+\bar{Q}(\bar{\eta})\approx \bar{Q}(\bar{\eta})
\end{eqnarray}
and appropriate boundary conditions. Here, the boundary terms $Q(\eta)$ and $\bar{Q}(\bar{\eta})$ are necessary because the variation of the differential equations $G_a$ and $\bar{G}_a$ leaves some boundary terms in general. We demand these terms to be cancelled by $\delta Q(\eta)$ and $\delta\bar{Q}(\bar{\eta})$. In the case at hand, these conditions are
\begin{eqnarray}
\delta Q(\eta)={k \over 2\pi}\int_{\partial\Sigma}\eta_a\delta A^a,\,\,\,\,\,
\delta \bar{Q}(\bar{\eta})=-{\bar{k} \over 2\pi}\int_{\partial\Sigma}\bar{\eta}_a\delta \bar{A}^a
\end{eqnarray}
Since $G(\eta)$ and $\bar{G}(\bar{\eta})$ are now differentiable, one can calculate easily their Poisson algebra,
\begin{eqnarray}
\{G(\eta),\,G(\lambda)\}&=&\int_{\Sigma}[\eta,\lambda]^aG_a-{k \over 2\pi}\int_{\partial\Sigma}\eta^a D\lambda_a,\nonumber\\
\{\bar{G}(\bar{\eta}),\,\bar{G}(\bar{\lambda})\}&=&\int_{\Sigma}[\bar{\eta},\bar{\lambda}]^a\bar{G}_a+{\bar{k} \over 2\pi}\int_{\partial\Sigma}\bar{\eta}^a D\bar{\lambda}_a
\end{eqnarray}

Therefore most of the structures in \cite{ban1} are retrieved in spite of the presence of the source term and GCST. 
Getting the black hole solution (\ref{bh}), we have used the gauge fixing condition $A^a_\rho=\bar{A}^a_\rho=0$, which is different from the one $\partial_\theta A^a_\rho=\partial_\theta \bar{A}^a_\rho=0$ used in \cite{ban1}. However, this difference does not alter those features of \cite{ban1} concerned with the gauge parameters significantly. One might think our condition completely fixes the gauge and leaves no residual symmetry. However, this is not the case. As is mentioned in the paper, the differential structure of the constraints inevitably results in the residual symmetry. In fact, if we require the gauge fixing condition remain the same under the time flow,   
\begin{eqnarray}
\partial_\tau A^a_\rho=D_\rho A^a_\tau=\partial_\rho A^a_\tau+[A_\rho,\,A_\tau]^a=\partial_\rho A^a_\tau=0,
\end{eqnarray}
thus the nondynamical gauge parameter $A^a_\tau$ is determined at most to be some unknown function of $\tau$ and $\theta$ and in the static case, function of $\theta$ only. The same equation also tells us that the condition $A^a_\rho=0$ is invariant under this residual symmetry, therefore makes consistency. Applying this analysis to the diffeomorphism with the gauge parameter $A^a_\tau=-\chi^i A^a_i=-\chi^\theta A^a_\theta$,  we conclude that the diffeomorphism parameter $\chi^\theta$ is a function of $\tau$ and $\theta$ and in the static case, $\theta$ only.

With this notion of the similarities with \cite{ban1}, one can straightforwardly obtain the affine algebra 
\begin{eqnarray}
\{T^a_n,\,T^b_m\}^*&=&-f^{ab}{}_cT^c_{n+m}-ikn\eta^{ab}\delta_{n+m},\nonumber\\
\{\bar{T}^a_n,\,\bar{T}^b_m\}^*&=&-f^{ab}{}_c\bar{T}^c_{n+m}+i\bar{k}n\eta^{ab}\delta_{n+m},
\end{eqnarray}
where $A^a_\theta=-{1 \over k}\sum_{n=-\infty}^{\infty}{T^a_n e^{in\theta}}$ and $\bar{A}^a_\theta={1 \over \bar{k}}\sum_{n=-\infty}^{\infty}{\bar{T}^a_ne^{in\theta}}$. As for the diffeomorphism, we make use of the beautiful result that the diffeomorphism with the parameter $\xi^i$ is equivalent, on shell,  to  the gauge transformation with the field dependent parameters $\eta^a=\xi^iA^a_i$ and $\bar{\eta}^a=-\bar{\xi}^i\bar{A}^a_i$\cite{witten}. In fact, this relation is realized as the Sugawara construction.
\begin{eqnarray}
J[\xi]={k \over 4\pi}\int \xi(\theta)A^a_\theta A_{a\theta}=\sum{L_n\xi^n},\,\,\,\,\,
\bar{J}[\bar{\xi}]={\bar{k} \over 4\pi}\int \bar{\xi}(\theta)\bar{A}^a_\theta \bar{A}_{a\theta}
=\sum{\bar{L}_n\bar{\xi}^n},
\end{eqnarray}
where $\xi^n={1 \over 4\pi}\int d\theta\xi(\theta)e^{in\theta}$ and $\bar{\xi}^n={1 \over 4\pi}\int d\theta\bar{\xi}(\theta)e^{in\theta}$. The Fourier modes $L_n={1 \over 2k}\sum{T_mT_{n-m}}$ and $\bar{L}_n={1 \over 2\bar{k}}\sum{\bar{T}_m\bar{T}_{n-m}}$ satisfy the classical Virasoro algebras with vanishing central charges. If we consider its quantization, we get the quantum central charges arising from the ordering ambiguity of the composite operators,

\begin{eqnarray}\label{quantum}
\hat{L}_n={2k \over 2k-Q}:L_n:,\,\,\,\,\,\hat{\bar{L}}_n={2\bar{k} \over 2\bar{k}+Q}:\bar{L}_n:,
\end{eqnarray}
where $Q\eta^{ad}\equiv f^a{}_{bc}f^{dbc}$. The results are

\begin{eqnarray}
c={2kN \over 2k-Q}={(l+\alpha)N \over l+\alpha+8G},\,\,\,\,\,\bar{c}={2\bar{k}\bar{N} \over 2\bar{k}+Q}={(l-\alpha)\bar{N} \over l-\alpha-8G},
\end{eqnarray}
where $N$ and $\bar{N}$ are the dimensions of the corresponding Lie algebras, i. e., the right moving part $sl(2,\Re)$ and the left moving part $sl(2,\Re)$.

In particular, the zero modes are related with the Casimir invariants as

\begin{eqnarray}
L_0={l \over 2(l+\alpha)}(Ml+J),\,\,\,\,\,\bar{L}_0={l \over 2(l-\alpha)}(Ml-J).
\end{eqnarray}
Their quantum counterparts (\ref{quantum}) are given by
\begin{eqnarray}
\hat{L}_0={l(Ml+J) \over 2(l+\alpha+8G)}={k^2(R_++R_-)^2 \over 2(k+1)l^2},\,\,\,\,\,\hat{\bar{L}}_0={l(Ml-J) \over 2(l-\alpha-8G)}={\bar{k}^2(R_+-R_-)^2 \over 2(\bar{k}-1)l^2}.
\end{eqnarray}
For $k={l+\alpha \over 8G}\gg1$ and $\bar{k}={l-\alpha \over 8G}\gg1$, one can use Cardy's formula to calculate the statistical entropy for the black hole.
\begin{eqnarray}\label{entropy}
S&=&2\pi\sqrt{{\Delta c \over 6}}+2\pi\sqrt{{\bar{\Delta}\bar{c} \over 6}}\nonumber\\  
&=&{\pi k\sqrt{k} \over 1+k }{R_++R_- \over l}+{\pi\bar{k}\sqrt{\bar{k}}  \over 1-\bar{k} }{R_+-R_- \over l}\end{eqnarray}
Here, $\Delta$ and $\bar{\Delta}$ denotes the eigenvalues corresponding to $\hat{L}_0$ and $\hat{\bar{L}}_0$ respectively. In the second equality of the equation, we used $N=\bar{N}=3$ and (\ref{mj}). Although the above statistical entropy is proportional to the area  in the $\alpha\rightarrow0$ limit, its coefficient is not the one expected from the area law. The correct value should come in the order $O(k,\bar{k})$, rather than the above $O(1/\sqrt{k} ,1/\sqrt{\bar{k}})$ term. This fact together with that the entropy (\ref{entropy}) is not proportional to the area when $\alpha$ is considerably large suggests that it cannot be the leading order of the entropy. If we demand the correct area law, the leading terms of the central charges should be

\begin{eqnarray}\label{central}
c=12{({l \over 8G})^2(k+1) \over k^2},\,\,\,\,\,\bar{c}=12{({l \over 8G})^2(\bar{k}-1) \over \bar{k}^2},
\end{eqnarray}
which reduce to $c=\bar{c}=3l/2G$ in the large $k$ and vanishing $\alpha$ limit. However, these values (\ref{central}) cannot be obtained from the present context. In the following section we discuss about the present status of BTZ black hole in regard to the statistical entropy and give some possible directions to go.

\section{Conclusions}

The basic observations we have made so far are as follow. 1. The BTZ black hole solution can be extended to the case with the GCST. This might not be possible in the Einstein gravity. If we impose the torsion free condition {\it a priori}, we will get the discrepancy in the derivative order between the Einstein-Hilbert term and the GCST. This makes the theory dynamical and the Einstein-Hilbert term dominates only in the low momentum limit, i.e., in the asymptotic region. Therefore, only in the asymptotic region, the BTZ geometry can be taken over. This asymptotic BTZ geometry is presumably not extended to the whole region. (If it is, the system is meant to be non-dynamical. One has to remember the Poincar\'{e} case where the asymptotic structure of the conical space and helical time cannot be extended to the whole region in the Einstein context \cite{clement}.) 

2. The BTZ black hole spectrum is shown to be deformed as the GCST is turned on. This deformation can be summarized as the boosting plus rescaling of the `vector' $(Ml,\, J)$ (see eq. (\ref{MJ})). In analogous to the Poincar\'{e} case, the gravitational mass and angular momentum are different from their inherent values. BTZ black hole itself can be understood as the adS$_{2+1}$ boosted in a specific way, compactified along the spatial isometric direction and rescaled so that the whole procedure is well defined in the extremal limit. (see the eq. (\ref{decompose}).) The effect of the GCST is to give further boosting, therefore to give more angular momentum that is realized as the induced spin. In this sense, the GCST gives effectively the description of the system in the rotating frame. This might have relevance in the DLCQ reduction of adS$_{2+1}$ into adS$_{1+1}$ discussed in \cite{strominger2}. In fact, $\alpha\rightarrow l$ limit reduces automatically the gauge group $SO(2,2)$ of the action (\ref{polar}) into $SO(2,1)$. This will be dealt with in more detail in other place \cite{cho1}. 

3. In the context of gauge gravity, the identification Killing vector of \cite{ban} is realized as the Wilson loop operators. In the original paper \cite{ban}, they classified in full detail all possible one parameter subgroups corresponding to the Killing vectors, of  which one is selected to make the black hole. This selection becomes natural in this gauge gravity context because it amounts to encode the nontrivial holonomy of the source onto the `target' manifold. 

4. The generic boundary WZW theory becomes chiral in the presence of the GCST. It assigns different Kac-Moody levels and different Virasoro central charges to the right moving part and the left moving part. In fact, this chirality breaking supports the interpretation of the GCST as to give the description in the rotating frame. This rotation of the frame in one direction breaks the chiral matching between the right moving sector and the left moving sector. 

5. The statistical entropy does not come in the leading order in $k$ and $\bar{k}$. This means that the central charge is too small to account for the are law precisely. The central charge depends on the gauge fixing condition. In this paper, we have chosen $A^a_\rho=\bar{A}^a_\rho=0$ as the gauge fixing condition. In the case, there is no classical contribution to the central charge, while the normal ordering ambiguity in the Sugawara construction brings quantum contribution to the central charge. In \cite{ban1}, the author fixed the gauge as $\partial_\theta A^a_\rho=\partial_\theta\bar{A}^a_\rho=0$ so that  there are classical contributions to the central charges like $c=12k(A_\rho)^2=12k(b^{-1}\partial_\rho b)^2$ and $\bar{c}=12\bar{k}(\bar{A}_\rho)^2=12\bar{k}(\bar{b}^{-1}\partial_\rho\bar{b})^2$, where $b(\rho)$ and $\bar{b}(\rho)$ are group elements. Those two gauge fixing conditions are therefore related to each other via gauge transformation $b(\rho)$ and $\bar{b}(\rho)$. 

Does the gauge transformation change the physics? Definitely it should not. In fact, there is a subtlety here; the constant elements $b^{-1}\partial_\rho b\equiv\beta$ and $\bar{b}^{-1}\partial_\rho\bar{b}\equiv\bar{\beta}$ imply that $b(\rho)=\exp{\beta\rho}$ and $\bar{b}(\rho)=\exp{\bar{\beta}\rho}$ are not well defined over the whole region of the `world' manifold. Therefore this singular gauge transform might relate two different vacuum sectors. For careful analysis, one has to distinguish those two sets of gauge connection fields. The gauge connection components of \cite{ban1} were obtained directly from the nondegenerate `dreibein' fields $e^a{}_\mu$ of the `target' manifold geometry. Therefore they are the gauge connections on the `target' manifold. On the other hand, those of this paper are defined on the `world' manifold. As is mentioned before, the components $e^a{}_\mu$ in this paper cannot compose the `dreibein' set because they are degenerate. The true soldering forms are defined as the covariant derivatives of the isovectors and this set of soldering forms gives the pull back mapping of the `target' geometry onto the `world' manifold. Consequently, the sigular gauge transformations $b(\rho)$ and $\bar{b}(\rho)$ relate the trivial solution of this paper corresponding to the `unbroken phase' and the nontrivial solution of \cite{ban1} corresponding to the `broken phase' \cite{witten}. 

Anyway it is interesting to see that although the bare central charge of our result is different from that of \cite{ban1}, the effective central charge entering into the Cardy's formula to count the degeneracy is always fixed and independent of $b$'s \cite{carlip}. Since the classical central charge does not contribute to the effective central charge, the statistical entropy is not the correct order in $G$, that is, we are still left with the black hole entropy problem in $(2+1)$-dimension. (The problems of recent approaches to BTZ black hole entropy are well described in \cite{carlip}.) In order to increase the central charge $c$ ($\bar{c}$), the multiplicity in the entropy counting through the Cardy' formula, we need to find more gauge symmetries concerned with the boundary CFT. All these difficulties originate from our ignorance about the way to specify {\it a priori} the precise boundary condition for the black hole asymptotics. This job is rather easy on the horizon; one can give apparent horizon condition as in \cite{carlip1}. As for the asymptotic boundary, one usually read off the boundary condition {\it a posteriori}  from the black hole solution. In most cases, these boundary conditions are determined to the leading order in $1/r$ expansion, so we are not sure whether it is black hole asymptotics or of other smooth matter distribution. Therefore it is crucial to our understanding of black hole entropy to determine precise boundary condition for the black hole asymptotics; once it is done, we will get the correct boundary theory with sufficient gauge symmetries to give the effective central charge $c_{eff}=3l/2G$.

\section*{Acknowledgement}
The author thanks T. Lee and S.-J. Rey for helpful discussions and comments. He is supported by the Korea Science and Engineering Foundation (KOSEF) through CTP. He is also indebted to Asia Pacific Center for Theoretical Physics (APCTP) for the hospitality during his visit.

\end{document}